\documentclass[10pt]{iopart}
\usepackage{textcomp}
\usepackage{graphicx}
\usepackage{bm}
\usepackage{upgreek}
\usepackage{IEEEtrantools}
\usepackage{multirow}
\usepackage[colorlinks=true,allcolors=blue,urlcolor=blue]{hyperref}
\usepackage{txfonts}
\usepackage[normalem]{ulem} 
\usepackage{booktabs}
\usepackage{longtable}
\usepackage{natbib}
\usepackage{aas_macros}

\newcommand{\eqn}[1]{Eq.~\ref{#1}}
\newcommand{\sect}[1]{Sect.~\ref{#1}}
\newcommand{\fig}[1]{Fig.~\ref{#1}}
\newcommand{\tab}[1]{Table~\ref{#1}}

\begin{document}

\title[Targeted optimization in small-scale atomic structure calculations]{Targeted optimization in small-scale atomic structure calculations: application to Au I}
\author{Sema~Caliskan,
Jon~Grumer, and
Anish~M.~Amarsi}

\address{Theoretical Astrophysics, Department of Physics and Astronomy, Uppsala University, Box 516, SE-751 20 Uppsala, Sweden}
\ead{sema.caliskan@physics.uu.se}
\vspace{10pt}
\begin{indented}
\item[]November 2023
\end{indented}

\begin{abstract}

The lack of reliable atomic data can be a severe limitation in astrophysical modelling, in particular of events such as kilonovae that require information on all neutron-capture elements across a wide range of ionization stages. Notably, the presence of non-orthonormalities between electron orbitals representing configurations that are close in energy can introduce significant inaccuracies in computed energies and transition probabilities. Here, we propose an explicit targeted optimization method that can effectively circumvent this concern while retaining an orthonormal orbital basis set. We illustrate this method within the framework of small-scale atomic structure models of Au~I, using the \textsc{Grasp2018} multiconfigurational Dirac-Hartree-Fock atomic structure code. By comparing to conventional optimization schemes we show how a targeted optimization approach improves the energy level positioning and ordering. Targeted optimization also leads to better agreement with experimental data for the strongest E1 transitions.
This illustrates how small-scale models can be significantly improved with minor computational costs if orbital non-orthonormalities are considered carefully. These results should prove useful to multi-element atomic structure calculations in, for example, astrophysical opacity applications involving neutron-capture elements.
\end{abstract}

\vspace{2pc}
\noindent{\it Keywords}: atomic structure, atomic data, MCDHF calculations, numerical methods

\ioptwocol

 \section{Introduction} \label{sec:intro}

Recent advancements in astronomical spectral analysis have introduced new demands on atomic physics. A good example is the advent of the James Webb Space Telescope operating in the infrared spectral regime \citep{boker_2022} where transition data for most elements and their ions are sparse. Another important example is the observation of the first-ever kilonova following the neutron star merger (NSM) event GW170817 detected by LIGO/Virgo \citep{Abbott_2017}. This ground-breaking discovery revived the debate on the origin of the neutron-capture elements, suggesting that these mergers are major sources of r-process nucleosynthesis in the universe \citep[e.g.][]{Cowan_2021}. 

Reliable atomic data are increasingly required for accurate astrophysical modelling. For instance, when trying to identify individual spectral features, one needs complete atomic data and transition parameters with a high accuracy for each species considered. On the other hand, lower-accuracy data may suffice for modelling broad-band energy distributions and opacities; but this still requires completeness as well as some degree of accuracy, since too large uncertainties can prevent a detailed chemical analysis of the studied astrophysical material \citep{kasen_2017}.

Unfortunately, the current databases are both incomplete and of limited accuracy when it comes to some of the heavier r-process elements, in particular in the infrared spectral region \citep[e.g.][]{smartt_2017}. 
This lack of information on atomic energy levels and processes is partly due to the complexities involved in carrying out accurate atomic structure calculations for many of these elements.

A significant challenge present in many atomic systems is the presence of multiple configurations with similar energies built from the same set of subshells. This is particularly true in the lanthanide group of elements where the ground and first excited configurations have closely aligned 4f, 5d, 6s and 6p orbital energies, implying that different subshell occupations lead to varying screening effects, and thus ideally requiring a non-orthonormal representation of the orbitals used to build the various configurations. 

However, current state-of-the-art atomic physics codes focused on atomic structure, such as the configuration-interaction-based codes \textsc{Fac} \citep{gu_2008} and \textsc{Hullac} \citep{hullac_2001}, assume a fully orthonormal orbital basis set. The current version of the multiconfigurational Dirac-Hartree-Fock (MCDHF) code \textsc{Grasp} \footnote{For the latest development version and further information and documentation of \textsc{Grasp} and associated codes, see also the open-source git repository at GitHub: \url{https://github.com/compas}.}
\citep{froese_fischer_2019,jonsson_2023} has, since the \textsc{Grasp2k} release \citep{Jonsson_2007}, allowed for some degree of non-orthonormality through biorthogonal transformations between even and odd parity orbital sets \citep{malmqvist_1986}, i.e. we obtain a different set of orbitals for each parity. Nevertheless, performing separate \textsc{Grasp}  calculations of two competing configurations of the same parity may reveal significant non-orthonormalities between the orbitals optimized on the two different configurations. These non-orthonormalities, together with the possible strong configuration interaction among these states, can cause inaccurate expectation values, such as energies and transition rates, as well as wrong level ordering. The latter may be a particular issue for obtaining a grossly realistic representation of e.g. resonance lines and thermodynamics in plasma modelling (e.g. in kilonova modelling; \citealt{pognan_2023}).

In this work, within the framework of the latest development version of the \textsc{Grasp} code (\textsc{Grasp}2018), we discuss a method for conducting atomic structure calculations that address the issue of orbital non-orthonormalities, while also ensuring that the wavefunctions remain compact enough for efficient computations of complete spectroscopic, large-scale opacity, or electron-scattering calculations that can be applied to a wide range of elements. 
The method, referred to as ``explicit targeted optimization (TO)'', consists of circumventing the orthonormality condition of \textsc{Grasp} by carefully optimizing the orbitals on targeted eigenstates. We also investigate semi-empirical rescalings of diagonal matrix elements to further improve the quality of the wavefunction, as recently discussed in the context of the \textsc{Grasp} code by \cite{li_2023}. 

We apply our methods to neutral gold, which is a representative system for illustrating some of the challenges of the lanthanides, while having the advantage of having a simpler atomic structure. Although the atom is a nominal one-electron system with a $5d^{10} 6s$ ground configuration, gold is a heavy, highly relativistic element that displays competing states similar to the lanthanides. This makes neutral gold a challenging system for theoretical computations \citep{gaarde_1994}. 

On the experimental side, several works on the spectrum of gold have been reported. The measurements of energy levels and lines made by \cite{platt_1941} were updated by the works of \cite{ehrhardt_1971} and \cite{brown_1978}. The latter two are the references used by the current version of NIST database \citep{NIST_ASD}, which will be used in this paper to compare our theoretical results to experimental values. \cite{george_1988} also revised the energy values of a few levels. More recently, \cite{civis_2011} measured several high-lying levels and revised a few of the NIST values. The availability of experimental energy levels and transition probabilities is another good reason to use gold to test our theoretical methods. 

On the theoretical side, however, calculations for gold in the literature are limited in number, or focused on only a few states or transitions (see \citealt{migdalek_2020} and references therein). \citet{safronova_2004} present a set of calculations for the $5d^{10} nl$ Rydberg series of Au~I using third-order many-body perturbation theory. They obtained good agreement with experimental data as compiled by the NIST database. However, discrepancies were found for some $5d^{10} nl$ states that were attributed to important interactions with the omitted $5d^{9} nl n'l'$ states.  More recently, \citet{mccann_2022} calculated energies for the lowest 26 levels of Au~I using the \textsc{Grasp}$^0$ \citep{parpia_grasp92_1996} and \textsc{Fac} \citep{gu_2008} codes, where they obtained an average difference of 8.6\% between their \textsc{Grasp}$^0$ energies and experimental data.

In the following, we present how our targeted optimization calculations with \textsc{Grasp2018} are done and how we apply them to the case of gold in \sect{sec:theory}--\ref{sec:method}. In \sect{sec:results}, we discuss the results obtained using our method, by presenting the calculated energy levels as well as transition probabilities for Au I, and comparing them with available Au~I data in the literature. Finally, we summarize our findings and discuss possible future work in \sect{sec:concl}.

\section{Atomic structure methods}
\label{sec:theory}

\subsection{Theoretical background}

\subsubsection{\textsc{Grasp}}

The theoretical investigations in this work are mainly performed with the \textsc{Grasp2018} suite of codes \citep{froese_fischer_2019,jonsson_2023b,jonsson_2023} making use of the MCDHF and relativistic configuration interaction (RCI) methods. A \textsc{Grasp} eigenstate is represented by an atomic state function (ASF; $\Psi$) that is defined by an expansion of configuration state functions (CSFs; $\Phi$): 
\begin{IEEEeqnarray}{rCl}
\label{eq:wavefunction}
    \Psi(\Gamma, P, J, M_J) & = & \sum_{i=1}^{N_{CSF}} c_i \Phi(\gamma_i, P, J, M_J),
\end{IEEEeqnarray}
where $c_i$ are the mixing coefficients, $J$ and $M_J$ are the angular quantum moments, $P$ is parity and $\gamma_i$ represents the occupied subshells of the CSF with their complete angular coupling tree information. The $\Gamma$ label of the ASF is usually taken from the dominant CSF component for the specific eigenstate solution. The CSF basis is constructed through substitutions -- singles (S), doubles (D), triples (T) etc. -- from a set of reference configurations within a specified set of active orbitals. Expanding the CSF basis essentially takes care of electron correlation effects. A multi-reference can also be adopted to include higher-order correlation effects in e.g. a SD model.

In the MCDHF scheme of \textsc{Grasp}, radial orbitals are obtained through a Dirac-Coulomb self-consistent-field (SCF) procedure where each electron moves in the field of the remaining electrons, and the radial functions are varied to minimize a weighted linear combination of a selected set of eigenenergies in the so-called extended optimal level (EOL) scheme. The SCF procedure is continued until the orbitals have converged. Based on this orbital basis set, subsequent RCI calculations may be employed to add relativistic corrections beyond the Dirac-Coulomb approximation, in the form of the Breit interaction and quantum electrodynamical effects, to obtain the final wavefunctions for the targeted states \citep{jonsson_2023b}.

It is well-known that the structure of neutral atoms are challenging to compute accurately, due to the relatively large screening effects among the outer electrons in comparison to the nuclear potential, i.e. large electron correlation effects. Moreover, when doing multiconfigurational calculations, balancing the correlation corrections among the targeted eigenstates is challenging, and the wavefunction representing each individual state may be less accurate than in a single-configuration calculation. Due to these difficulties faced with neutrals and multiconfigurational calculations, convergence issues often occur with the MCDHF SCF procedure in \textsc{Grasp}. To overcome this, we used the DBSR-HF code \citep{zatsarinny_2016} to obtain better initial estimates of the orbitals.

\subsubsection{\textsc{Fac}}

To complement the atomic structure investigation carried out with \textsc{Grasp}, we also perform RCI calculations with the \textsc{Fac} code \citep{gu_2008}. While there are many similarities between the two approaches, e.g. in how the atomic eigenfunctions are represented in terms of $jj$-coupled CSF's  (\eqn{eq:wavefunction}) and the use of a Dirac-Coulomb Hamiltonian at the core of both codes, a fundamental difference between the two implementations is the representation of the electron-electron interactions and how the set of radial orbitals is optimized. While \textsc{Grasp} is using a MCDHF-EOL scheme to optimize the orbitals,
\textsc{Fac} instead employs the computationally more efficient and more practical Dirac-Fock-Slater method. In this approach, a common, local central potential is optimized on an average configuration that is explicitly specified or determined from a set of selected configurations. This means that, in \textsc{Fac}, all orbitals are defined by a common central potential, while in  \textsc{Grasp}, the MCDHF approach allows for a larger degree of freedom with the sacrifice of a more complex computational procedure. As mentioned in the introduction, \textsc{Grasp} also makes use of biorthogonal transformation techniques to allow for separate treatment of the even and odd parity states, i.e. including some level of non-orthogonality among the orbitals, while \textsc{Fac} is restricted to use a common orbital set for all states. See the \textsc{Fac} git repository for further information and documentation of the code\footnote{The \textsc{Fac} open-source repository: \url{https://github.com/flexible-atomic-code/fac}.}.

\subsection{Spectroscopic CI models\label{sec:SCI}}

We compare the atomic structure results obtained with the explicit targeted optimization method using \textsc{Grasp} (described in \sect{sec:method}) to more conventional spectroscopic CI models (SCI) with \textsc{Grasp} (MCDHF + RCI) and \textsc{Fac} (RCI), i.e. models with a CI space limited to include only those CSF's that represent physical states. SCI atomic structure models are of particular current interest as they are the predominant foundation in the atomic data used for computations of atomic processes and, ultimately, multi-element modelling of r-process dominated  kilonovae plasmas \citep[e.g.][]{kasen_2017,fontes_2020,tanaka_2020,pognan_2023}.

In all cases considered in this work, we focus on the first four even and two odd excited configurations (see \tab{tab:configs}). In the case of \textsc{Grasp}, the orbitals are optimized on the even and odd states in two separate MCDHF calculations. For \textsc{Fac}, typical calculations use a central potential optimized on the ground configuration as a starting point. For the present case of Au~I, however, the central potential was instead optimized on the average configuration formed by 5d$^{10}$6s and 5d$^{9}$6s$^2$ for a better energy level structure in comparison to experimental energy levels.

\section{The computational challenges for gold}
\label{sec:challenges}

\begin{table*}
\caption{The first few energy levels of Au~I calculated using three different orbital optimization schemes, and experimental values from the NIST database (in cm$^{-1}$). ``Opt. $5d^9$ + $5d^{10}$'' indicates that the orbitals are optimized on both the $ 5d^9$ and the $ 5d^{10}$ states. The other 2 columns indicate whether they are optimized on the $ 5d^9$ or $ 5d^{10}$ states. The root-mean-square deviation is calculated by taking the difference between the calculated and experimental energies. \label{tab:energies1}}
\begin{tabular}{c c c c c c c c c c}
\toprule
 & \multicolumn{2}{c}{Main component} & & &  &   &  &  \\
 \cmidrule{2-3}
 Level index & Conf. & Term & J  & $E$(Opt. $5d^9$+$5d^{10}$) & $E$(Opt. $5d^{10}$) & $E$(Opt. $5d^{9}$) & $E$(Exp.) \\ [0.5ex] 
 \hline
1 & $5d^{10}6s$ & $^2S$ & 1/2      & 0.0& 0.0  &   0.0    &  0.0\\ 
2 & $5d^{9}6s^{2}$ & $^2D$ & 5/2   & 7216.7 & 14 306.9  &  2457.8  & 9161.2 \\
3 & $5d^{9}6s^{2}$ & $^2D$ & 3/2   & 19 193.5 & 26 901.5 & 14 000.4 & 21 435.2 \\
4 & $5d^{10}6p$ & $^2P^o$ & 1/2  & 35 051.2  & 30 257.9 & 30 809.4 & 37 359.0    \\
5 & $5d^{10}6p$ & $^2P^o$ & 3/2   & 36 563.5  & 32 876.6 & 32 225.8 & 41 174.6 \\
6 & $5d_{5/2}^{9}6s_{1/2} 6p_{1/2}$ & $^4P^o$ & 5/2  &29 473.0 & 38 748.5 & 24 412.1 & 42 163.5        \\
7 & $5d_{5/2}^{9}6s_{1/2} 6p_{1/2}$ & $^4F^o$ & 7/2  & 33 098.0 & 41 827.6 &28 059.3 & 45 537.2  \\
8 & $5d_{5/2}^{9}6s_{1/2} 6p_{1/2}$ & $^4F^o$ & 5/2 & 34 173.7  & 43 260.2  & 29 115.2 &   46 175.0 \\
9 & $5d_{5/2}^{9}6s_{1/2} 6p_{3/2}$ & $^4D^o$ & 5/2  & 40 077.9 & 48 784.4  & 35 020.9 &   46 379.0        \\
10 & $5d_{5/2}^{9}6s_{1/2} 6p_{3/2}$ & $^4P^o$ & 3/2 & 34 690.1 & 43 897.9  & 29 629.0 &   47 007.4       \\
11 & $5d_{5/2}^{9}6s_{1/2} 6p_{3/2}$ & $^4F^o$ & 9/2 & 35 270.8 & 44 435.1 & 30 197.4  &   48 697.2      \\
12 & $5d_{5/2}^{9}6s_{1/2} 6p_{3/2}$ & $^4D^o$ & 7/2 & 38 926.7 & 47 570.1 & 33 878.0   &  51 028.9        \\
13$^a$ & $5d_{5/2}^{9}6s_{1/2} 6p_{1/2}$ & $^2D^{o}$ & 3/2 & 43 835.3 & 48 894.1            & 38 751.2   &  51 231.5       \\
14 & $5d_{5/2}^{9}6s_{1/2} 6p_{3/2}$ & $^4F^o$ & 3/2 &  39 613.1 & 53 568.7 & 34 548.5 &  51 485.0      \\
15 & $5d_{5/2}^{9}6s_{1/2} 6p_{3/2}$ & $^2D^o$ & 5/2 &  51 458.2 & 59 817.2 & 46 396.4  &  51 653.9     \\
16 & $5d_{5/2}^{9}6s_{1/2} 6p_{3/2}$ & $^2F^o$ & 7/2 &  50 209.2 & 57 873.0  & 45 161.1   &  52 802.1      \\
17 & $5d_{5/2}^{9}6s_{1/2} 6p_{3/2}$ & $^4P^o$ & 1/2 &  40 441.1 &57 770.7 & 35 361.9  &  53 196.3     \\
18 & $5d^{10}7s$ & $^2S$ & 1/2  & 40 568.0 & 42 098.6  & 38 558.4  &   54 485.2       \\
19$^b$ & $5d_{3/2}^{9}6s_{1/2}6p_{1/2}$ & $^4D^{o}$ & 1/2 & 47 653.7 &  57 770.7    & 42 537.0 &   55 732.5      \\
20$^c$ & $5d_{3/2}^{9}6s_{1/2}6p_{1/2}$ & $^4D^o$ & 3/2  & 49 604.2 &  60 079.2   & 44 522.5 &   56 105.7        \\
21$^d$ & $5d_{3/2}^{9}6s_{1/2}6p_{1/2}$ & $^2F^{o}$ & 5/2   & 46 736.8 & 56 568.4   & 41 629.0   &58 616.8       \\
22 & $5d_{3/2}^{9}6s_{1/2}6p_{1/2}$ & $^2P^o$ & 3/2    & 50 277.4 & 59 679.1 & 45 166.4 &  58 845.4      \\
23 & $5d_{3/2}^{9}6s_{1/2}6p_{3/2}$ & $^2F^o$ & 5/2    & 66 440.5  & 72 705.2 & 61 562.3 &  59 713.2   \\
24 & $5d_{3/2}^{9}6s_{1/2}6p_{3/2}$ & $^2D^o$ & 5/2    &  57 886.1 & 64 004.8 & 53 042.8 &  61 255.1    \\
25 & $5d_{3/2}^{9}6s_{1/2}6p_{3/2}$ & $^2D^o$ & 3/2    & 69 247.9 & 74 911.5 & 64 402.5 &   61 563.3   \\
26 & $5d^{10}6d$ &  $^2D$ & 3/2  & 46 343.9  & 48 457.3 & 44 631.1     &61 951.6        \\
27 & $5d^{10}6d$ &  $^2D$ & 5/2 & 46 397.7  & 48 509.0 &  44 686.9 & 62 033.7  \\
28 & $5d_{3/2}^{9}6s_{1/2}6p_{3/2}$ & $^2P^o$ & 3/2  & 55 757.3  & 59 679.1  & 51 055.5   &   63 005.1   \\
29 & $5d_{3/2}^{9}6s_{1/2}6p_{3/2}$ & $^2P^o$ & 1/2   & 67 133.1  & 72 737.0   & 62 316.3 &  $65\,566.3^e$   \\
\cmidrule{1-8}
\multicolumn{4}{c}{root-mean-square deviation} & 9382.6	& 6793.8& 13 024.9 &  \\
\br
\end{tabular}
\newline $^a$In our calculations, this term is highly mixed with the $^4F_{3/2}$ term which is the dominant LSJ term
\newline $^b$In our calculations, this term is highly mixed with the $^4P_{1/2}$ term which is the dominant LSJ term
\newline $^c$In our calculations, this term is highly mixed with the $^2P_{3/2}$ term which is the dominant LSJ term
\newline $^d$In our calculations, this term is highly mixed with the $^4F_{5/2}$ term which is the dominant LSJ term
\newline $^e$From \citet{civis_2011}
\end{table*}

As mentioned in the introduction, non-orthonormalities between the orbitals can be an important aspect of the computation of heavy r-process elements, such as the lanthanides. The electronic valence shell structure of neutral lanthanides can be summarised as $4f^{q_1} \, 5d^{q_2} \, 6s^{q_3} \, 6p^{q_4}$, where the $q$'s represent the electron occupation of each configuration. Depending on the subshell occupation, the shape of the 4f, 5d, 6s, and 6p orbitals may vary significantly.
Therefore, in a small-scale spectroscopic calculation of a wide range of energy eigenstates of a given lanthanide system, say, it would often be a much too crude approximation to assume that e.g. a single 4f orbital is representative of all the 4f orbitals in the various configurations, and a non-orthonormal orbital set should ideally be used.

Moreover, the states belonging to  different competing configurations are sometimes nearly degenerate and thus potentially highly mixed. 
The mixing coefficients for the configuration interactions and the relative energy positions of the different interacting configuration groups are therefore sensitive to the representation of their respective orbitals and can be challenging to compute correctly with an orthonormal orbital set \citep[e.g.][]{carlsson_1988}.

Their complex shell structure and large number of levels make the lanthanides a difficult test-bench for model development. A simpler case with similar challenges in terms of non-orthonormalities between orbitals can be found in the copper-group elements. These elements do not have open f-shells and thus have considerably fewer levels per configuration. Their atoms have $nd^{10} \, n'l$ Rydberg series accompanied by a number of bound states belonging to the configurations $nd^{9} \, n's^2$ and  $nd^{9} \, n's \, n'p$, with n = 3 to 5 and n'= 4 to 6 across the homologous group. Similarly for the lanthanides, these configurations involve the same s, p, and d orbitals with different electron occupations. Moreover, the $nd^{9}$ states can interact and perturb the $nd^{10}$ states \citep{carlsson_1988,gaarde_1994}.

In our work, we considered neutral gold, which is the third element in the copper-group, as a test case for the explicit targeted optimization method. Gold is an interesting element from an astrophysical point of view, in the sense that its cosmic origin is unknown to this day \citep[e.g.][]{Kobayashi_2020}. It lies close to the third r-process peak \citep[e.g.][]{Roederer_2022}, and like most of the lanthanides, it might be produced in NSM's \cite[e.g.][]{gillanders_2021}.

\begin{figure}
\centering
\includegraphics[width=\hsize]{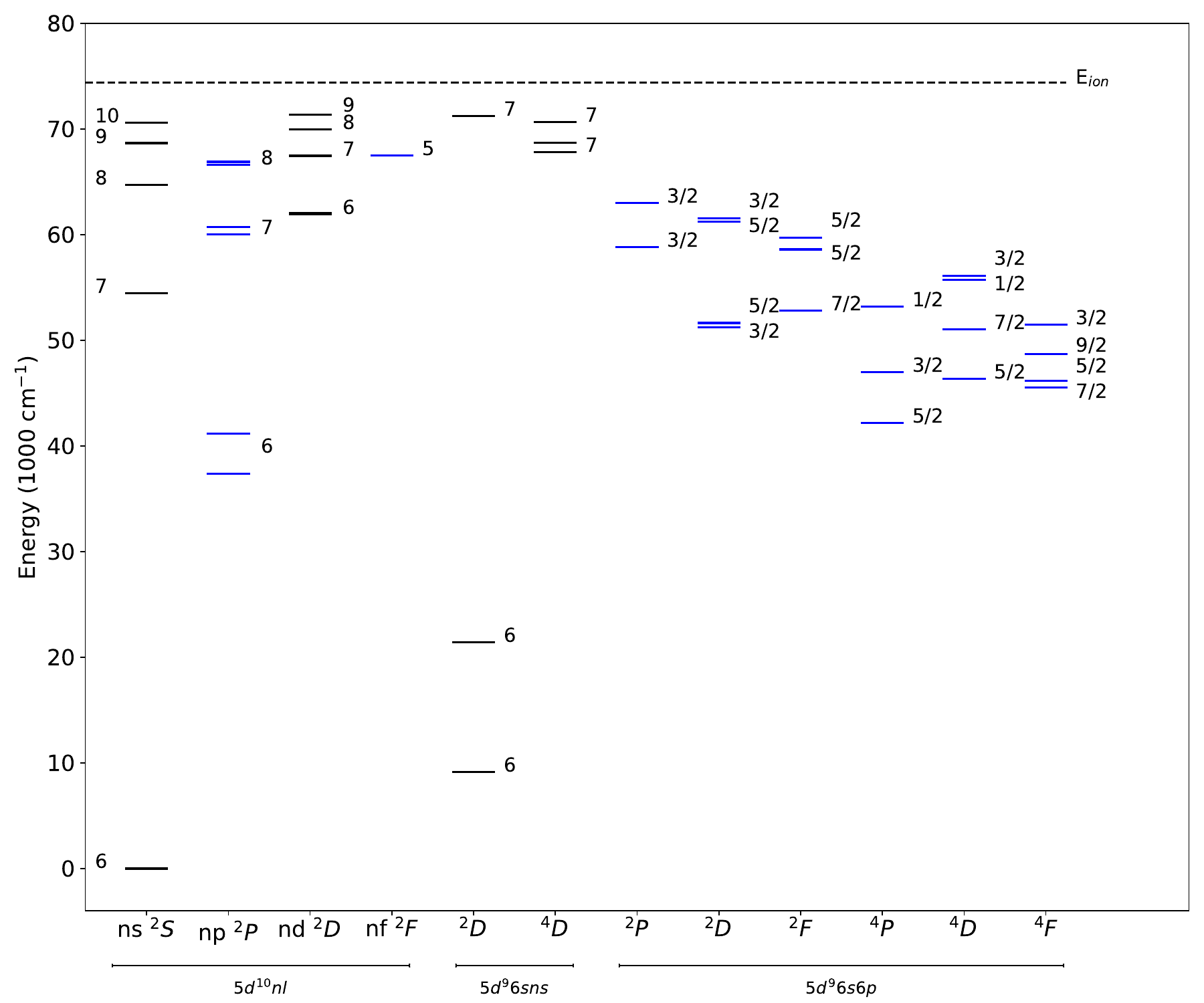}
\caption{Partial energy-level diagram for neutral gold. The black lines represent the even parity states while the blue ones represent the odd parity levels. The numbers next to the levels of Rydberg series represent the principal quantum number n. For the $5d^9 6s 6p$ states, the fractions are the values of J. The energy values are experimental values from the NIST database.
\label{fig:term}}
\end{figure}

The electronic structure of the gold atom can be seen in the partial $LS$-term energy-level diagram in \fig{fig:term}, which can help to better understand the possible interactions between the different states. For example, we pointed out above that the $5d^9 6s^2$ and $5d^{9} 6s 6p$ may interact with states in the $5d^{10} nl$ Rydberg series. However, we can see from \fig{fig:term} that the $5d^9 6s^2$ $^2D_{3/2, 5/2}$ terms, which can only interact with the $5d^{10} 6d$ $^2D_{3/2, 5/2}$ states, are situated far below those. The levels belonging to the $5d^{9} 6s 6p$ configuration, on the other hand, interlap with the $5d^{10} np$ and $5d^{10} nf$ series, and since they have the same parity, a strong interaction between these states is expected, perturbing the Rydberg series.

\begin{figure}
\centering
\includegraphics[width=\hsize]{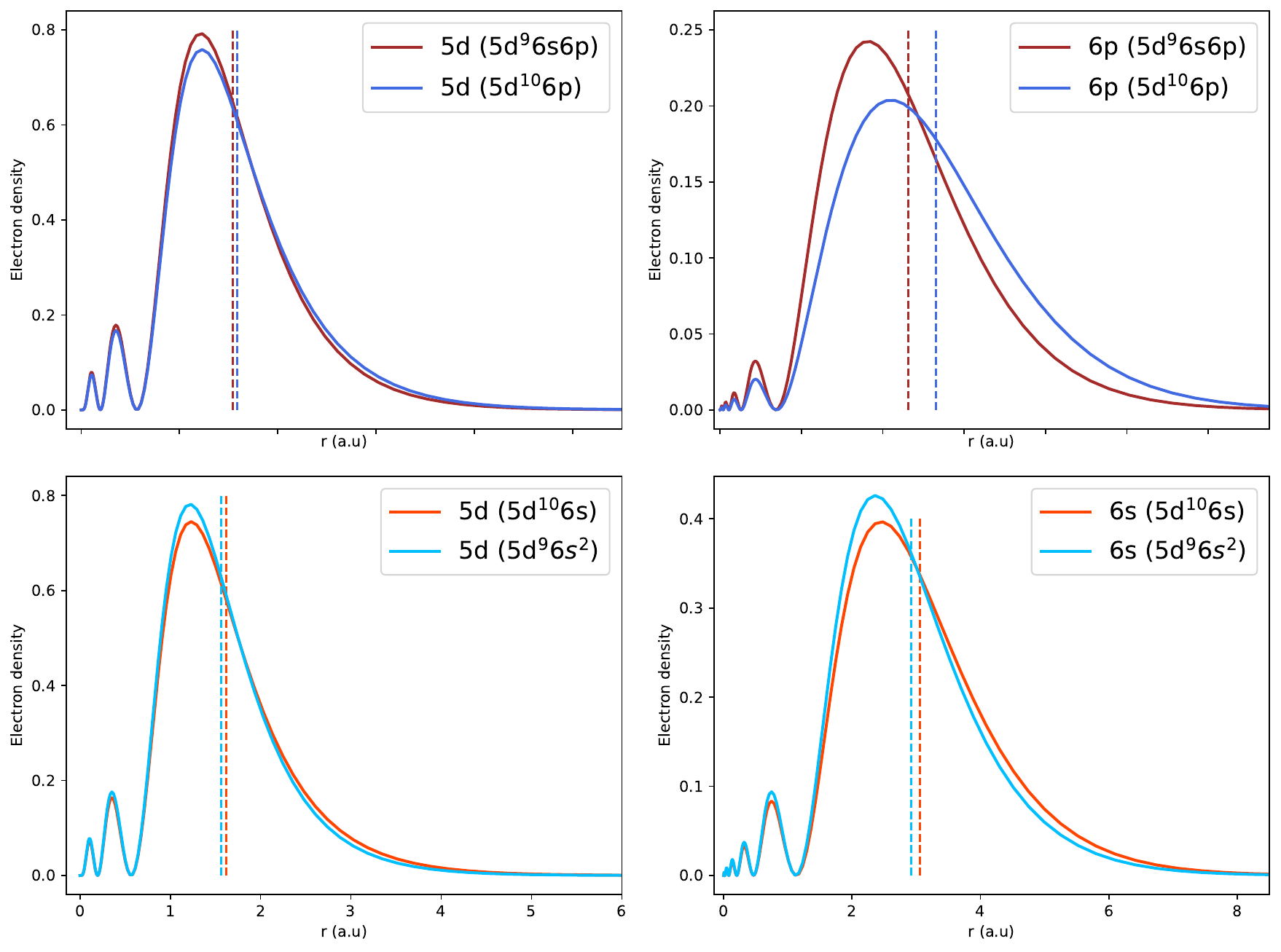}
\caption{The electron density representing the shape of the 5d, 6s and 6p orbitals optimized for different configurations, based on separate Dirac-Fock calculations. The dashed vertical lines represent the $\langle r\rangle$ expectation values.
\label{fig:density}}
\end{figure}

On top of the perturbed series, another computational difficulty is the shape of the radial orbitals. In \fig{fig:density}, it can be seen how the shape of the 5d, 6s and 6p orbitals vary from states with ten 5d-electrons to states with nine 5d-electrons. To further illustrate the differences between them, the expectation values of their radii are represented with a dashed vertical line, and the values are given in \tab{tab:radius}. The 6s and 6p valence orbitals are more extended when optimized for the $5d^{10}$ states, since they feel more screening effects due to the additional tenth 5d electron. This variation in optimal distribution of the radial one-electron functions is most severe for the 6p orbital, optimized for either the $5d^{10} 6p$ or $5d^{9} 6s 6p$ states, as also shown by the difference in the expectation values of the radius of 6p in \tab{tab:radius}. The computed orbitals that are ideal for representing these various configurations are therefore non-orthonormal.

The combination of strong configuration mixing and non-orthonormalities between the orbitals has a big impact on the computed energy levels of Au I. In \tab{tab:energies1} (fifth column) and on \fig{fig:energies}, we show the energies resulting from the conventional \textsc{Grasp} SCI model, described in \sect{sec:SCI}. We can see that when the orbitals are forced to be orthonormal (in the sense that the different configurations are computed together using a common orbital set), the level ordering between the $5d^{10} 6p$ and $5d^{9} 6s 6p$ states is wrong. 

Moreover, to illustrate the sensitivity of the calculated energies to the way that the orbitals are optimized, we also show energies from the SCI model but where the orbitals are optimized either on only the $5d^9$ or only on the $5d^{10}$ states separately. The results are shown in \tab{tab:energies1} and \fig{fig:energies}. We observe that when optimized on the $5d^{10}$ states, the levels belonging to the $5d^{10} nl$ configurations get pushed down in energy, thus the $5d^{9}$ states appear higher in the energy level structure as compared to experiment. The opposite occurs when we optimize them on the $5d^{9}$ states. The MCDHF-SCI calculation with orbitals optimized on all states ($5d^{9}$+$5d^{10}$) seems to be an in-between situation among the other two extremes, but still gives the wrong energy ordering compared to the experimental energy levels. It should be pointed out that separate Dirac-Fock-type calculations with \textsc{Grasp}, for each configuration, would not be optimal either, since this would omit important configuration interactions.

One could improve the SCI energies with a large-scale correlation model. The most dominant correlation contributors to the calculated energies are often the configurations belonging to the Layzer complex \citep{froese_fischer_1997}. By adding these to the reference set of configurations, a multi-reference is created that captures interactions arising from single and double substitutions, as well as the dominant higher-order interactions from triple and quadruple substitutions in the correlation model. However, for systems such as Au I, this implies a multi-reference with configurations such as $5d^8 5f^2 nl$ and $5d^7 5f^2 nl^2$ that represent the important $d^2 \rightarrow{} f^2$ substitution or substitutions to 5g (although in practice these are likely less important). Large-scale multiconfigurational calculations with such a multireference would quickly become computationally very expensive, especially when the final wavefunctions are to be used to compute other properties, e.g. radiative and electron-impact rates, or opacity-oriented calculations across a wide range of elements and their ions. 

In the next section, we address the issue of finding a balance between computational cost, achieved accuracy and applicability, through an explicit targeted optimization approach.

\begin{table}
\caption{\label{tab:radius} Expectation values of the radius $\langle r\rangle$, in atomic units, of the 5d$^-$, 5d, 6s, 6p and 6p$^-$ orbitals optimized on the states belonging to the configurations in the first row, based on separate Dirac-Fock
calculations.}
\begin{indented}
\item[]\begin{tabular}{lcccc}
\hline
$\langle r\rangle$ 
& $5d^{10}6s$ & $5d^{9}6s^2$ & $5d^{10}6p$ & $5d^9 6s 6p$  \smallskip \\
\hline 
5d & 1.62 & 1.56     & 1.59 & 1.54 \\
5d$^-$ & 1.54 & 1.49 & 1.52 & 1.48 \\
6s & 3.06 & 2.92 & -    & 2.80 \\
6p & -   & -         & 5.30 & 4.63 \\
6p$^-$ & -     & -    & 4.75 & 4.10 \\
\hline
\end{tabular}
\end{indented}
\end{table}

\section{Explicit targeted optimization}
\label{sec:method}

We showed in the previous section how Au~I demonstrates clear non-orthonormal effects in the orbital basis. We also remind ourselves from the introduction that \textsc{Grasp} and most other current, state-of-the-art atomic structure codes are limited to using orthonormal orbital basis sets. Given this constraint, to correct for the differences between the optimal shape of some of the orbitals belonging to the $5d^{10}\,nl$ and $5d^9nl\,n'l'$ states in Au I, and without expanding the model to a large-scale correlation calculation, we investigate a so-called explicit targeted optimization method approach. We focus our attention on low-energy configurations of Au~I.

To construct a TO model for Au I, we start by noting that the 5d, 6s, and 6p orbitals show non-orthonormal effects. To remedy this while retaining a compact correlation model, we add a few carefully chosen configurations to the CI space, built from a few new orbitals that are specifically tailored to address the non-orthonormality issues. 

In more detail, we start from orbitals generated in a $5d^9 6s\, nl$ MCDHF calculation (with $nl$ being $6s$ or $6p$ for the even and odd parity calculations, respectively). Then, to represent the remaining $5d^{10}nl$ states (which, as we learned in the previous section, ideally should be described by more extended $5d$ orbitals), we allow for some radial variation in the subshell by expanding the original configuration as $5d^{10}nl = 5d^{9}(5d + n'd)nl = 5d^{10}nl, 5d^{9}nl \, n'd$ and adding these two configurations into the configuration basis, where the $n'd$ orbital is a new, ``correction orbital'' with the same symmetry as the original $5d$. For the present range of configurations included in the SCI model, the next available orbital to use for this purpose in the even calculation is $7d$, and $6d$ for the odd.

Furthermore, to correct also for the larger non-orthonormalities present in $6s$ and $6p$ orbitals, we expand the configurations as $5d^{9}(5d + n'd)(nl+n'l)$, where $n'l$ are correction orbitals with $s$ or $p$ symmetries for the even and odd calculation, respectively. Expanding this final expression results in a set of new configurations, see \tab{tab:configs}, corresponding to a slightly enlarged CSF space. Using this basis, the final important step is then to optimize the correction orbitals on the $5d^{10}nl$ states only, to force them to correct for the non-orthonormalities with the orbitals from the initial $5d^9 6s\, nl$ calculation. Since the correction orbitals are optimised within the active set containing all the other orbitals, the orthonormality is automatically ensured during the MCDHF optimisation procedure through the use of Lagrange multipliers, so that an orthonormal set of orbitals is obtained both for the even and odd parities separately.

The procedure can be summarized as follows:
\begin{enumerate}
    \item For each parity, identify non-orthonormalities and group configurations accordingly.
    \item MCDHF calculation of $5d^{9}6s\,nl$ states.
    \item Express $5d^{10}nl$ as $5d^{9}(5d + n'd)(nl+n'l)$.
    \item Explicitly add correction configurations to the CSF basis.
    \item MCDHF optimization of correction orbitals on $5d^{10}nl$ states.
    \item Perform final RCI calculation to obtain all ASF eigenstates. 
\end{enumerate}

The scheme outlined here for Au~I is readily generalized to any atomic system with similar challenges. After identifying non-orthonormalites, the configurations are grouped with those that have approximately orthonormal ideal orbitals. An important aspect to pay attention to in the construction of a TO model, is which configuration (or configuration group) to start from. In the case of Au I, we started from the $5d^{9}$ states, but one could also perform step (ii) on the $5d^{10}$nl group. This, however, has the disadvantage of producing correction configurations with $5d^8$ subshells (due to the expansion of $5d^{9}6s\,nl$, see step (iii)), resulting in a substantially more complex CSF space. Moreover, this will potentially cause a larger CSF space if a correlation calculation based on S and D substitutions from the correction configurations is adopted, due to the presence of configurations with $5d^6$ subshells.

The difference between the TO approach and a more traditional MCDHF model lies in the initial observation of present orbital non-orthonormalities, and based on this, how the CSF basis is constructed and the configuration-specific orbital optimization scheme. This procedure of explicitly adding configurations follows what is sometimes called a ``configuration-driven approach'', in contrast to the more commonly applied ``orbital-driven approach'' \citep[e.g.][]{jonsson_2023b}, where the CSF basis is constructed through substitutions within an active set of orbitals.

Another important distinction of the TO method is how we selectively optimize the orbitals on the states of different configurations to coerce them to correct for the non-orthonormalities that we know are present. This is in contrast to conventional approaches where the orbitals usually are optimized on all states.

\begin{table}
\caption{\label{tab:configs} The list of the spectroscopic configurations and the correction configurations used in the targeted optimization calculation.}
\begin{indented}
\item[]\begin{tabular}{c|cc}
\hline
& even parity & odd parity \smallskip \\\hline
spectroscopic configurations & $5d^{10}6s$ & $5d^{10}6p$ \\
& $5d^{9}6s^2$ & $5d^{9}6s6p$ \\
& $5d^{10}7s$  &              \\
& $5d^{10}6d$  &              \smallskip \\\hline
correction configurations & $5d^{10}8s$ & $5d^{10}7p$ \\
& $5d^{9}6s7d$ & $5d^{9}6p6d$ \\
\hline
\end{tabular}
\end{indented}
\end{table}

\section{Results and Discussion} 
\label{sec:results}

\subsection{Energy levels}

\begin{table*}
\caption{The first few energy levels of Au~I calculated using the explicit targeted optimization method (with TO) and without the TO method, as well as with \textsc{Fac}, literature values from \citet{mccann_2022}, and experimental values from the NIST database (in cm$^{-1}$). The last two lines are theoretically predicted energy levels for which no experimental could be found in the literature. The root-mean-square deviation is calculated by taking the difference between the calculated and experimental energies. \label{tab:energies2}}
\begin{tabular}{c c c c c c c c c c}
\toprule
 & \multicolumn{2}{c}{Main component} & & &  &   &  &  &  \\
 \cmidrule{2-3}
 Level index & Conf. & Term & J & $E$(Without TO) & $E$(With TO)  & $E$(\textsc{Fac}) & $E$(Lit.) & $E$(Exp.) \\ [0.5ex] 
 \hline
1 & $5d^{10}6s$ & $^2S$ & 1/2    & 0.0   & 0.0 & 0.0  & 0.0 &  0.0   \\ 
2 & $5d^{9}6s^{2}$ & $^2D$ & 5/2  & 13 365.4 &  13 259.0 & 14 007.0 &  6726.9 &              9 161.2    \\
3 & $5d^{9}6s^{2}$ & $^2D$ & 3/2  & 25 411.4  & 24 791.0 & 26 620.0 & 19 313.8  &             21 435.2 \\
4 & $5d^{10}6p$ & $^2P^o$ & 1/2   & 29 891.6 & 30 534.3  & 31 138.9 & 39 417.6  &            37 359.0  \\
5 & $5d^{10}6p$ & $^2P^o$ & 3/2   & 32 825.1  &33 691.1 & 33 444.3 & 40 910.1 &               41 174.6 \\
6 & $5d_{5/2}^{9}6s_{1/2} 6p_{1/2}$ & $^4P^o$ & 5/2 & 32 547.0 &34 793.1 & 33 222.6 & 40 218.7 &  42 163.5  \\
7 & $5d_{5/2}^{9}6s_{1/2} 6p_{1/2}$ & $^4F^o$ & 7/2 & 36 254.3 & 38 552.8 & 36 897.8 &  45 914.1 & 45 537.2 \\
8 & $5d_{5/2}^{9}6s_{1/2} 6p_{1/2}$ & $^4F^o$ & 5/2 & 37 344.7 & 39 619.6  & 38 118.6 & 46 693.2  & 46 175.0 \\
9 & $5d_{5/2}^{9}6s_{1/2} 6p_{3/2}$ & $^4D^o$ & 5/2 & 43 303.9 & 45 621.3  & 44 633.5 & 53 189.7  &    46 379.0  \\
10 & $5d_{5/2}^{9}6s_{1/2} 6p_{3/2}$ & $^4P^o$ & 3/2 & 37 818.7 & 40 086.5 &38 859.6& 46 715.2   &47 007.4\\
11 & $5d_{5/2}^{9}6s_{1/2} 6p_{3/2}$ & $^4F^o$ & 9/2 & 38 362.6 & 40 643.9 &39 961.1  & 47 999.1   &48 697.2 \\
12 & $5d_{5/2}^{9}6s_{1/2} 6p_{3/2}$ & $^4D^o$ & 7/2 & 42 209.6 & 44 534.4& 43 451.0  & 52 180.1   &     51 028.9              \\
13$^a$ & $5d_{5/2}^{9}6s_{1/2} 6p_{1/2}$ & $^2D^o$ & 3/2 &42 761.4  & 45 007.5& 44 128.7 & 51 982.6 &       51 231.5 \\
14 & $5d_{5/2}^{9}6s_{1/2} 6p_{3/2}$ & $^4F^o$ & 3/2 &  47 046.2    &  49 288.4&44 128.7  & 57 239.0  &   51 485.0\\
15 & $5d_{5/2}^{9}6s_{1/2} 6p_{3/2}$ & $^2D^o$ & 5/2 &  54 174.8 & 56 578.2  & 55 994.3 & 66 138.7   &     51 653.9               \\
16 & $5d_{5/2}^{9}6s_{1/2} 6p_{3/2}$ & $^2F^o$ & 7/2 &  52 626.2  &55 122.8 & 54 538.5 & 65 194.9    &52 802.1  \\
17 & $5d_{5/2}^{9}6s_{1/2} 6p_{3/2}$ & $^4P^o$ & 1/2 &  43 510.4  &45 740.4 & 45 304.2  & 53 540.8   &   53 196.3  \\
18 & $5d^{10}7s$ & $^2S$ & 1/2 & 43 673.4 & 49 187.2 &  46 774.6 & - & 54 485.2 \\
19 & $5d_{3/2}^{9}6s_{1/2}6p_{1/2}$ & $^4D^o$ & 1/2   & 52 646.5  & 54 572.0&52 776.8 & 62 221.1 &55 732.5        \\
20 & $5d_{3/2}^{9}6s_{1/2}6p_{1/2}$ & $^4D^o$ & 3/2   & 53 407.2  & 55 678.6 &54 356.0 &64 437.8   &  56 105.7       \\
21$^b$ & $5d_{3/2}^{9}6s_{1/2}6p_{1/2}$ & $^2F^{o}$ & 5/2 & 49 867.9 &52 112.1 & 51 323.3 &60 498.2    &    58 616.8                    \\
22 & $5d_{3/2}^{9}6s_{1/2}6p_{1/2}$ & $^2P^o$ & 3/2   & 52 711.6  & 54 899.9  &  55 564.2& 65 129.1  & 58 845.4      \\
23 & $5d_{3/2}^{9}6s_{1/2}6p_{3/2}$ & $^2F^o$ & 5/2   & 67 299.2 &  70 100.8 & 69 792.5 &      84 278.3  &   59 713.2   \\
24 & $5d_{3/2}^{9}6s_{1/2}6p_{3/2}$ & $^2D^o$ & 5/2   & 58 822.5  & 61 552.0  & 60 995.7 &74 259.2    &  61 255.1  \\
25 & $5d_{3/2}^{9}6s_{1/2}6p_{3/2}$ & $^2D^o$ & 3/2   & 69 607.8  & 72 237.1 & 72 410.1      & 86 242.6  & 61 563.3  \\
26 & $5d^{10}6d$ &  $^2D$ & 3/2  & 50 206.3     & 55 601.8 &53 386.2 & 65 041.3  &                    61 951.6 \\
27 & $5d^{10}6d$ &  $^2D$ & 5/2  & 50 256.5     & 55 657.6 &53 409.3 & 64 701.1 &                         62 033.7                  \\
28 & $5d_{3/2}^{9}6s_{1/2}6p_{3/2}$ & $^2P^o$ & 3/2  & 56 597.8  & 58 223.9  & 58 082.0    & -   &    63 005.1                \\
29 & $5d_{3/2}^{9}6s_{1/2}6p_{3/2}$ & $^2P^o$ & 1/2  & 69 607.8 & 69 511.2  & 70 423.3  & -  &     $65\,566.3^c$                 \\
 & $5d_{3/2}^{9}6s_{1/2}6p_{3/2}$ & $^4P^o$ & 1/2  & 50 787.4  &   52 999.4  & - & -  &  - \\
 & $5d_{5/2}^{9}6s_{1/2}6p_{3/2}$ & $^2F^o$ & 7/2  & 57 299.6 &    59 914.6  &   59 699.4 & -   &  - \\
\cmidrule{1-9}
\multicolumn{4}{c}{root-mean-square deviation} & 7418.0 & 5842.2 & 6721.8  & 8340.2 &  \\
 \hline
\end{tabular}
\newline $^{a}$In our calculations, this term is highly mixed with the $^4F_{3/2}$ term which is the dominant LSJ term: 0.47  $5d^{9}6s6p$  $^4F_{3/2}$ + 0.24 $5d^{9}6s6p$ $^2D_{3/2}$ (see level with index 13 in Table 3).
\newline $^{b}$In our calculations, this term is highly mixed with the $^4F_{5/2}$ term which is the dominant LSJ term: 0.41  $5d^{9}6s6p$  $^4F_{5/2}$ + 0.28 $5d^{9}6s6p$ $^2F_{5/2}$ (see level with index 21 in Table 3).
\newline $^c$From \citet{civis_2011}.
\end{table*}

\begin{table*}
\begin{center}
\caption{Theoretical LS coupling composition of the states of Au~I calculated with the explicit targeted optimization method, with level indices in the same ordering as in \tab{tab:energies1} and \tab{tab:energies2}. For clarity, we show only the three biggest contributors to the wavefunction. \label{tab:lscomp}}
\begin{tabular}{cll}
\toprule
Level index & State & $LS$-composition  \\
\midrule
1   & $5d^{10}\,6s~^{2}S_{1/2}$         & 0.95 + 0.02~$5d^{9}\,7d~^{2}S$ \\
2   & $5d^{9}\,6s^{2}~^{2}D_{5/2}$      & 1.00                         \\
3   & $5d^{9}\,6s^{2}~^{2}D_{3/2}$      & 1.00                          \\
4   & $5d^{10}\,6p~^{2}P_{1/2}^{\circ}$ & 0.90 + 0.05~$5d^{9}\,6s\,6p~^{2}P^{\circ}$                          \\
5   & $5d^{10}\,6p~^{2}P_{3/2}^{\circ}$ & 0.89 + 0.06~$5d^{9}\,6s\,6p~^{2}P^{\circ}$                        \\
6   & $5d^{9}\,6s\,6p~^{4}P_{5/2}^{\circ}$ & 0.83 + 0.14~$5d^{9}\,6s\,6p~^{4}D^{\circ}$ + 0.02~$5d^{9}\,6s\,6p~^{4}F^{\circ}$      \\
7   & $5d^{9}\,6s\,6p~^{4}F_{7/2}^{\circ}$ & 0.65 + 0.20~$5d^{9}\,6s\,6p~^{4}D^{\circ}$ + 0.09~$5d^{9}\,6s\,6p~^{2}F^{\circ}$        \\
8   & $5d^{9}\,6s\,6p~^{4}F_{5/2}^{\circ}$ & 0.51 + 0.23~$5d^{9}\,6s\,6p~^{2}F^{\circ}$ + 0.12~$5d^{9}\,6s6p~^{2}F^{\circ}$    \\
9  & $5d^{9}\,6s\,6p~^{4}D_{5/2}^{\circ}$ & 0.46 + 0.16~$5d^{9}\,6s\,6p~^{2}F^{\circ}$ + 0.12~$5d^{9}\,6s\,6p~^{2}D^{\circ}$   \\
10   & $5d^{9}\,6s\,6p~^{4}P_{3/2}^{\circ}$ & 0.57 + 0.22~$5d^{9}\,6s\,6p~^{4}D^{\circ}$ + 0.11~$5d^{9}~^{2}D\,\,6p~^{2}P^{\circ}$        \\
11   & $5d^{9}\,6s\,6p~^{4}F_{9/2}^{\circ}$ & 1.00 \\
12  & $5d^{9}\,6s\,6p~^{4}D_{7/2}^{\circ}$ & 0.75 + 0.10~$5d^{9}\,6s\,6p~^{2}F^{\circ}$ + 0.09~$5d^{9}\,6s\,6p~^{4}F^{\circ}$   \\
13  & $5d^{9}\,6s\,6p~^{4}F_{3/2}^{\circ}$ & 0.47 + 0.24~$5d^{9}\,6s\,6p~^{2}D^{\circ}$ + 0.17~$5d^{9}\,6s\,6p~^{2}D^{\circ}$        \\
14   & $5d^{9}\,6s\,6p~^{4}F_{3/2}^{\circ}$ & 0.46 + 0.19~$5d^{9}\,6s\,6p~^{4}P^{\circ}$ + 0.17~$5d^{9}\,6s\,6p~^{2}D^{\circ}$        \\
15  & $5d^{9}\,6s\,6p~^{2}D_{5/2}^{\circ}$ & 0.51 + 0.31~$5d^{9}\,6s\,6p~^{4}D^{\circ}$ + 0.08~$5d^{9}\,6s\,6p~^{2}D^{\circ}$   \\
16  & $5d^{9}\,6s\,6p~^{2}F_{7/2}^{\circ}$ & 0.68 + 0.22~$5d^{9}\,6s\,6p~^{4}F^{\circ}$ + 0.06~$5d^{9}\,6s\,6p~^{2}F^{\circ}$   \\
17  & $5d^{9}\,6s\,6p~^{4}P_{1/2}^{\circ}$ & 0.52 + 0.23~$5d^{9}\,6s\,6p~^{2}P^{\circ}$ + 0.19~$5d^{9}\,6s\,6p~^{4}D^{\circ}$        \\
18  & $5d^{10}\,7s~^{2}S_{1/2}$         & 1.00             \\
19  & $5d^{9}\,6s\,6p~^{4}D_{1/2}^{\circ}$ & 0.48 + 0.33~$5d^{9}\,6s\,6p~^{2}P^{\circ}$ + 0.16~$5d^{9}\,6s\,6p~^{2}P^{\circ}$   \\
20  & $5d^{9}\,6s\,6p~^{4}D_{3/2}^{\circ}$ & 0.45 + 0.15~$5d^{9}\,6s\,6p~^{4}P^{\circ}$ + 0.15~$5d^{9}\,6s\,6p~^{2}D^{\circ}$  \\
21  & $5d^{9}\,6s\,6p~^{4}F_{5/2}^{\circ}$ & 0.41 + 0.28~$5d^{9}\,6s\,6p~^{2}F^{\circ}$ + 0.11~$5d^{9}\,6s\,6p~^{2}D^{\circ}$        \\
22  & $5d^{9}\,6s\,6p~^{2}P_{3/2}^{\circ}$ & 0.64 + 0.16~$5d^{9}\,6s\,6p~^{4}D^{\circ}$ + 0.10~$5d^{9}\,6s\,6p~^{2}P^{\circ}$   \\
23  & $5d^{9}\,6s\,6p~^{2}F_{5/2}^{\circ}$ & 0.57 + 0.26~$5d^{9}\,6s\,6p~^{2}F^{\circ}$ + 0.10~$5d^{9}\,6s\,6p~^{2}D^{\circ}$  \\
24  & $5d^{9}\,6s\,6p~^{2}D_{5/2}^{\circ}$ & 0.61 + 0.18~$5d^{9}\,6s\,6p~^{2}D^{\circ}$ + 0.13~$5d^{9}\,6s\,6p~^{2}F^{\circ}$   \\
25  & $5d^{9}\,6s\,6p~^{2}D_{3/2}^{\circ}$ & 0.55 + 0.35~$5d^{9}\,6s\,6p~^{2}D^{\circ}$ + 0.04~$5d^{9}\,6s\,6p~^{2}P^{\circ}$   \\
26  & $5d^{10}\,6d~^{2}D_{3/2}$         & 1.00              \\
27  & $5d^{10}\,6d~^{2}D_{5/2}$         & 1.00              \\
28  & $5d^{9}\,6s\,6p~^{2}P_{3/2}^{\circ}$ & 0.73 + 0.10~$5d^{9}\,6s\,6p~^{2}P^{\circ}$ + 0.05~$5d^{10}\,6p~^{2}P^{\circ}$  \\
29  & $5d^{9}\,6s\,6p~^{2}P_{1/2}^{\circ}$ & 0.61 + 0.27~$5d^{9}\,6s\,6p~^{2}P^{\circ}$ + 0.08~$5d^{10}\,7p~^{2}P^{\circ}$ \\
30  & $5d^{9}\,6s\,6p~^{4}P_{1/2}^{\circ}$ & 0.46 + 0.32~$5d^{9}\,6s\,6p~^{4}D^{\circ}$ + 0.16~$5d^{9}\,6s\,6p~^{2}P^{\circ}$        \\
31  & $5d^{9}\,6s\,6p~^{2}F_{7/2}^{\circ}$ & 0.78 + 0.15~$5d^{9}\,6s\,6p~^{2}F^{\circ}$ + 0.04~$5d^{9}\,6s\,6p~^{4}F^{\circ}$   \\
\bottomrule
\end{tabular}
\end{center}
\end{table*}

\begin{figure}
\centering
\includegraphics[width=\hsize]{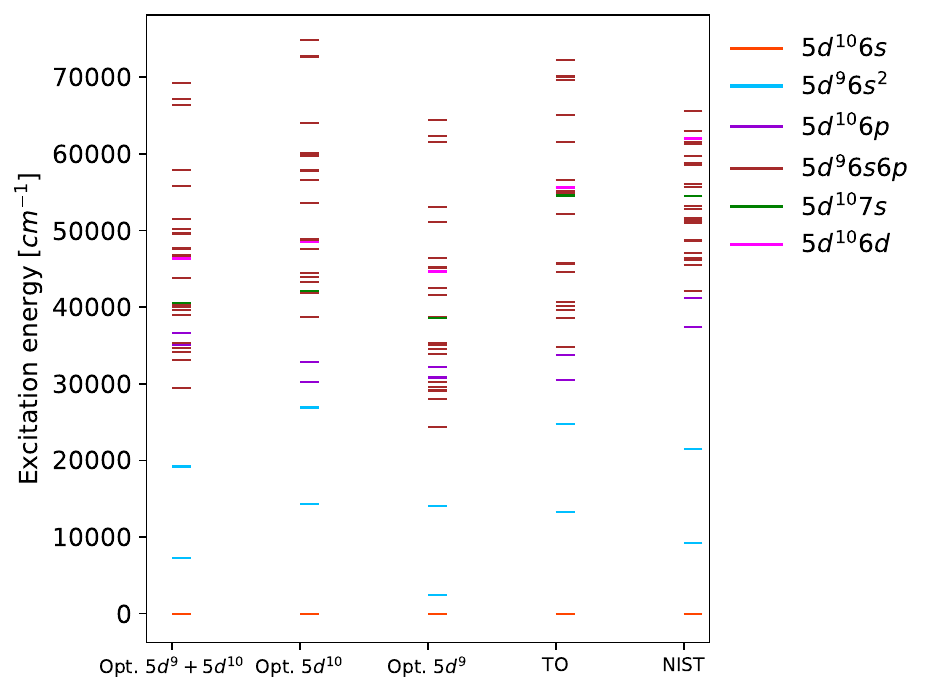}
\caption{Comparison between the calculated energy levels of Au~I (first four columns) and the experimental energies from NIST (last column). Each column represents a different optimization scheme:``Opt. $5d^9$ + $5d^{10}$'' is a spectroscopic calculation where the orbitals are optimized for all states, while the ``Opt. $5d^9$ or $5d^{10}$'' calculations are done by optimizing the orbitals either on the $5d^9$ states or $5d^{10}$ states. Finally, ``TO'' represents the results obtained with the targeted optimization method.
\label{fig:energies}}
\end{figure}

\subsubsection{TO versus other optimization schemes}

The energy values resulting from the explicit targeted optimization method are given in the sixth column of \tab{tab:energies2}. Compared to the more conventional MCDHF-RCI calculations of SCI type with the orbitals optimized together on all the $5d^{10}$ and $5d^9$ states (fifth column of \tab{tab:energies1}), we can see that the TO optimization corrects the wrong relative position of interacting states belonging to $5d^{10} 6p$ and $5d^{9} 6s 6p$, and improves the separation between the $5d^{10}6p$ $^2P_{1/2}$ and $^2P_{3/2}$ levels. Moreover, we observe a significant improvement in the even parity $5d^{10} nl$ states. The relative energy differences ($(E_{\mathrm{exp.}} - E_{\mathrm{calc.}})/E_{\mathrm{exp.}}$) drops from 25.5\% to 9.7\% for the $5d^{10} 7s$ $^2S_{1/2}$ level and from $\sim$ 25\%  to $\sim$ 10\% for the two $5d^{10} 6d$ levels. This shows the impact of the correction configurations on the $5d^{10}$ series. Overall, the energy results show that, when compared to the other orbital optimization schemes (i.e. optimized for $5d^{10}$ + $5d^9$ states simultaneously, or for $5d^{10}$ or $5d^9$; see  \sect{sec:challenges}), the TO method gives the best agreement with the experimental energies, as can also be seen in \fig{fig:energies}. 

Nevertheless, despite the overall improvement achieved by the TO method, we note that the first two excited levels $5d^{9} 6s^2$ $^2D_{5/2}$ and $^2D_{3/2}$ appear too high in energy compared to the experimental values and has large percentage differences with 44.7\% and 15.7\%, respectively. Remaining inaccuracies in the TO excitation energies are likely due to uneven energy contributions from the correction CSF's. For instance, since the spin-angular symmetry of the $5d^{10} 8s$ correction configuration is equal to the $5d^{10} 6s$ ground configuration, the eigenstates of the latter may be lowered more than the states with other symmetries, such as the ones belonging to $5d^{9} 6s^2$. Problems like this can be remedied by expanding the small-scale TO model and converging correlation contributions.

We also note that the calculated $5d^{9} 6s 6p$ energy levels are far too extended in energy, i.e. the configuration is too wide in energy compared to experiment. A possible explanation of such situations can be found in that low-spin states of a given configuration typically have higher excitation energies, and that these tend to require more electron correlation corrections than the lower-lying high-spin states, for which the electrons are forced to spatially avoid each other due to the Pauli principle. This often causes small-scale atomic structure models, i.e. models where correlation contributions are not converged, to predict configurations with too wide-spread excitation energy structures.

The improvement in the energy level structure from applying the TO method could potentially be attributed to the electron correlation contributions when we explicitly include correction configurations and thus slightly expand the CSF basis (\eqn{eq:wavefunction}). To exclude the effect of correlation in the comparison, we also compare our results with a calculation performed within the exact same CSF basis (i.e. the same spectroscopic and correction configurations), but without any targeted optimization in obtaining the orbitals. Instead, we first calculate the $5d^{10}$ and $5d^9$ states simultaneously, and then add the correction configurations and optimize the correction orbitals for the spectroscopic states; just as one would do for correlation orbitals.
The results from this calculation, labelled ``Without TO'', are shown in the fifth column of \tab{tab:energies2}. Comparing these results with the conventional SCI model ``Opt. $5d^{9}$+$5d^{10}$'', we can see that the impact of the competition between the $5d^{10} 6p$ and $5d^{9} 6s 6p$ states is less severe than in the SCI model, but the relative position of the levels is still improved significantly with TO. The representation of the even $5d^{10} nl$ states is also better when we do a targeted optimization, as well as the general agreement with the experimental values. This indicates once again that the optimization scheme of the orbitals has a significant impact on the calculated energies, and how a careful hands-on approach such as TO may improve the results significantly in small-scale calculations.

To further display the effect of the TO method on the energy structure as a whole, we also estimated the root-mean-square deviation for the energies obtained with the different optimizations schemes (see bottom of \tab{tab:energies1} and \tab{tab:energies2}). We find that the deviation from the experimental energies is the smallest for the TO method compared to the other methods.

However, it should also be noted that Au~I represent an atomic system that is neither well-represented in either jj- or LSJ-coupling. This is particularly true for the $5d^{9} 6s 6p$ states, as can be seen in \tab{tab:lscomp}. The strong mixing for some states complicates their labelling and therefore also the level-by-level matching with the experimental levels. Detailed analysis of the dominant LS terms has been done to identify the levels, but some possible differences between the theoretical and experimental dominant LS terms have been reported in the captions of \tab{tab:energies1} and \tab{tab:energies2}. These differences could be due to the theoretical model we use and the representation of the wavefunction, or due to the labelling of the experimental levels. Nevertheless, one should be cautious when considering the differences between the theoretical and experimental energies, since this could in some cases be caused by the high mixing in the $5d^{9} 6s 6p$ states.

\subsubsection{Comparison of TO to \textsc{Fac} SCI and other works}

To further test the TO results, we also compare to \textsc{Fac} SCI calculations (described in \sect{sec:SCI}). The resulting energies are reported in   \tab{tab:energies2}. The obtained excitation structure is comparable to the TO results in terms of relative energy differences to experimental results (an absolute average of 11.5\% with TO versus 13.7\% with \textsc{Fac}). However, the relative energy positions of the $5d^{10}6p$  $^2P_{3/2}$ and $5d^{9}6s6p$  $^4P_{5/2}$ states remain an issue with the \textsc{Fac} SCI model.

Interestingly, the \textsc{Fac} SCI model gives energies that are in better agreement with experimental values than energies obtained with the corresponding \textsc{Grasp} SCI ``Opt. $5d^{9}+5d^{10}$'' model (with an absolute average difference of 16.2\%). One should note that the \textsc{Fac} model was tuned to obtain the optimal central potential yielding the most accurate energy level structure, which partly explains the good agreement with TO results and experiments. In contrast, the TO model relies only on observing the non-orthonormalities through separate smaller calculations on the various configurations, and then taking these differences into account via state-selective optimization of correction orbitals.

\begin{table*}[h!]
\begin{center}
\caption{Transition rates ($s^{-1}$) calculated with different models, compared to experimental values from NIST, for selected E1 transitions of Au~I. \label{tab:rates}}
\begin{tabular}{c c c c c c c c c c c}
\hline
\multirow{1}{*}{$\lambda_{\mathrm{air}}$} & 
\multirow{2}{*}{Lower} &
\multirow{2}{*}{Upper} & 
\multirow{2}{*}{Acc.} & 
\multirow{2}{*}{$A$(NIST)} & 
\multirow{2}{*}{$A$(TO)} & 
\multirow{1}{*}{$ \% $}  & 
\multirow{1}{*}{$A$(Without} & 
\multirow{1}{*}{$\%$} & 
\multirow{1}{*}{$A$(Opt.} &
\multirow{1}{*}{$\%$} \\
(nm) & & & & & & diff. & (TO) & diff. & $5d^9 + 5d^{10}$) & diff. 
\smallskip \\
\hline
242.8 & $5d^{10}6s$ $^2S_{1/2}$  & $5d^{10}6p$ $^2P_{3/2}$     & $\leq 7 \%$ & 1.98E+08&	2.01E+08 & 1.4 &	1.71E+08 & $-13.7$ &	1.68E+08 &	$-14.9$ \\
267.6 & $5d^{10}6s$ $^2S_{1/2}$ & $ 5d^{10}6p$ $^2P_{1/2}$	& $\leq 2 \%$ &    1.64E+08	&   1.63E+08 & $-0.6$ &  1.37E+08 &	$-16.5$ &  1.75E+08 & 6.6 \\
274.8 & $5d^{9}6s^2$ $^2D_{5/2}$ & $5d^9 6s6p$  $^4F_{7/2}$& $\leq 10 \%$ & 1.03E+07	&  4.49E+06	 & $-56.4$ & 3.29E+06 & $-68.0$ & 3.90E+06 &	$-62.1$ \\
312.3 & $5d^{9}6s^2$ $^2D_{5/2}$ & $5d^{10} 6p$ $^2P_{3/2}$	& $\leq 7 \%$&    1.90E+07	 &5.25E+06	& $-72.4$ & 5.44E+06 & $-71.4$ & 2.95E+07 & 55.2 \\
406.5 & $5d^{10} 6p$ $^2P_{1/2}$ & $5d^{10} 6d$ $^2D_{3/2}$& $\leq 7 \%$	&  8.50E+07 &	1.16E+08 & 37.1 & 4.36E+07 & $-48.8$ &	7.78E+06 &	$-90.8$ \\
479.2 & $5d^{10} 6p$ $^2P_{3/2}$ & $5d^{10} 6d$ $^2D_{5/2}$&   $\leq 7 \%$&  8.90E+07 &	1.32E+08 & 47.8  & 5.12E+07 & $-42.5$ &	9.10E+06 &	$-89.8$ \\
481.2 & $5d^{10} 6p$ $^2P_{3/2}$ & $5d^{10} 6d$ $^2D_{3/2}$& $\leq 7 \%$	&  1.57E+07 &	2.23E+07 & 42.2 & 8.69E+06 & $-44.7$ &	1.53E+06 &	$-90.3$ \\
583.7 & $5d^{10} 6p$ $^2P_{1/2}$ & $5d^{10} 7s$ $^2S_{1/2}$	& $\leq 7 \%$ &   2.95E+07	&   4.22E+07 &  42.9 & 1.13E+07	  & $-61.7$ &5.80E+05 &	 $-98.0$ \\
751.1 & $5d^{10} 6p$ $^2P_{3/2}$ & $5d^{10} 7s$ $^2S_{1/2}$ & $\leq 7 \%$	 & 4.24E+07	 &  7.40E+07	& 74.5 & 1.93E+07 & $-54.4$ & 7.83E+05 &	$-98.2$ \\
 \hline
\end{tabular}
\end{center}
\end{table*}

We also compared our calculated energies with the most recent theoretical calculation of these levels in the literature, done by \citet{mccann_2022}. The authors reported energy levels for the $5d^{10}6s$, $5d^{9}6s^2$, $5d^{10}6d$, $5d^{10}6p$, and $5d^{9}6s6p$ configurations, calculated using a modified version of the  \textsc{Grasp}$^0$ code \citep{parpia_grasp92_1996}. They included correlation by explicitly adding important even and odd parity configurations, resulting in a multireference of a total of 14 configurations, giving rise to 2202 CSFs. Their choice of configurations was done with the motivation to achieve a comparatively good representation of the system while keeping the calculation compact enough for a subsequent R-matrix electron-impact calculation.

It is important to point out that the size of the calculation, i.e. the number of CSFs available to the wavefunction representation, is an important parameter to take into account. Indeed, a larger CSF expansion typically captures more correlation contributions. However, we need to find a balance between the accurate representation of the wavefunction and the size of the calculation. This is particularly difficult for heavy systems such as gold, where the number of CSFs can quickly grow to unmanageable proportions, as was discussed in the end of \sect{sec:challenges}. Considering that, while the \textsc{Grasp}$^0$ calculation of \citet{mccann_2022} is relatively compact for being a modern \textsc{Grasp} calculation, the TO model achieves a comparable accuracy in the energy level structure with an order of magnitude smaller basis size (with 146 CSFs in total). The TO model also has the benefit of being a systematic and fully ab-initio approach, based only on theoretical observations. Moreover, we can see the issue of competing levels belonging to $5d^{10}6p$ and $5d^{9}6s6p$ in their data set, which is successfully addressed with the TO method. 

\subsection{Transition properties}

Despite several theoretical and experimental works on the transitions of Au~I \citep{jannitti_1979,hannaford_1981,gaarde_1994,migdalek_2000,safronova_2004,fivet_2006,zhang_2018}), only 20 lines out of the 191 lines listed in the NIST database have an available transition rate to date. The quasi-totality of these lines lies in the ultraviolet and visible regions, and only very few lines have been observed in astrophysical objects. In the Sun, the most useful Au~I line is the $312.28$ nm (wavelength in air)
between $5d^{9} 6s^{2}\,{^2D_{5/2}}$ and $5d^{10} 6p\,{^2P^{\mathrm{o}}_{3/2}}$ \citep{Grevesse_2015}, with a precise transition rate measured by \citet{hannaford_1981} using laser-induced excitation.  In metal-poor stars, the $267.59$ nm line ($5d^{10} 6s\,{^2S_{1/2}}$ and $5d^{10} 6p\,{^2P^{\mathrm{o}}_{1/2}}$) can be used as an abundance indicator \citep{Cowan_2002,Sneden_2003,Roederer_2012,Placco_2015}, while the $242.79$ nm ($5d^{10} 6s\,{^2S_{1/2}}$ and $5d^{10} 6p\,{^2P^{\mathrm{o}}_{3/2}}$) can sometimes also be detected \citep{Roederer_2022}.  \citet{gillanders_2021} have recently searched for a number of Au~I lines in the AT2017gfo kilonova, placing upper limits on the abundance of gold in the ejecta.  

In order to evaluate the impact of the TO method on the transition rates, we computed E1 transitions for Au~I based on the three calculation models: TO, without TO, and MCDHF SCI with both $5d^9$ and $5d^{10}$ states. The resulting A-values and experimental A-values from NIST are given in \tab{tab:rates}. We selected transitions with A-values higher than $1.0\times 10^{7}$ $s^{-1}$, since weak transitions are more challenging to compute. We also selected the ones that have a NIST accuracy classification of at least lower than 10$\%$.  This selection contains three stellar lines mentioned above, as well as one of the lines considered by \citet{gillanders_2021}, namely the $751.07$ nm (air). We only present the A-values with the Babushkin gauge (or length form), which is the preferred gauge for transitions involving lowly excited states \citep{hibbert_1974}, while there is an indication that the Coulomb gauge is more suitable for transitions with high-lying Rydberg states \citep{papoulia_2019}.

\tab{tab:rates} also shows the difference in percentage between the calculated and experimental transition rates, to allow for a better comparison between the different optimization schemes. We once again observe that the optimization scheme of the orbital basis have a significant impact on the calculated properties, like the transition rates. When comparing to the experimental A-values, the TO method gives the best agreement among the three tested calculation models, improving the transition rates by up to 20\% for some transitions. Notably, the experimental A-values of the two strongest transitions (at 242.8 and 267.6 nm), which are also the diagnostics for gold in stellar atmospheres, agree within the experimental uncertainty with the values from TO, whereas the other optimization schemes do not. The A-values of transitions involving the $5d^{9} 6s^2$ $^2D_{3/2}$ and $^2D_{5/2}$ states have the poorest agreement with NIST values, probably due to the important difference between the calculated and experimental energies.

\subsection{Fine-tuning of energies}

Theoretical energy levels and transition properties can potentially be further improved without increasing the size of the CSF basis by semi-empirically fine-tuning the diagonal elements of the Hamiltonian matrix to observed excitation energies. Adjustment of ab-initio diagonal energy parameters to  experimental energies has been used extensively in CI calculations by \citet[e.g.][]{hibbert_2003}, where it was demonstrated that the fine-tuning process gives a better description of the mixing between different LS terms with a common J-value than ab-initio calculations. It has also been shown that fine-tuning can circumvent the problem of wrong relative positions of interacting states and perturbers within Rydberg series \citep{carlsson_1988, lundberg_2001}. 

Fine-tuning has so far been applied to LSJ-coupled multiconfigurational Hartree–Fock (MCHF) and configuration interaction (CI) calculations. However, it was not applied to jj-coupled multiconfigurational Dirac-Hartree–Fock (MCDHF) or relativistic configuration interaction (RCI) calculations due to the typical presence of large off-diagonal Hamiltonian matrix elements. In their work, \cite{li_2023} investigated a method that can overcome this problem. In their proposed method, the Hamiltonian matrix is transformed from jj to LSJ coupling, applying fine-tuning in this representation, and then transforming back to jj-coupling again, which ultimately modifies both the diagonal and off-diagonal elements. This method has been implemented in the development version of \textsc{Grasp} through the new programs \texttt{jj2lsj\_2022} and \texttt{rfinetune}.

We applied this fine-tuning procedure to the Au~I energy matrix obtained with the TO method and we obtained mitigated results. For the even parity levels, the computed energies rapidly converged to experimental values as the \textit{jj-LSJ-jj} fine-tuning method was applied iteratively. However, this was not the case for the odd-parity levels, which diverged more and more at each iteration of the fine-tuning process. This could be explained by the fact that the $5d^{9} 6s 6p$ odd parity configuration is subject to large off-diagonal matrix elements in either the jj or LSJ representation, and the best theoretical diagonal energies are therefore far from the experimental ones.
A possible way forward could be to apply a non-linear fit of the eigenvalues by varying the diagonal matrix elements.

\section{Conclusions} 
\label{sec:concl}
In this paper, we investigated the application of a method within the realm of MCDHF calculations using the \textsc{Grasp}2018 atomic structure code to improve theoretical excitation energies and transition rates through a small-scale calculation using a compact basis set, such as those employed in multi-element astrophysical opacity calculations. This was achieved by a targeted orbital optimization approach focusing on the perturbing configurations of neutral gold, in order to address the issue of non-orthonormality between the orbitals. 

The efficacy of the TO method was assessed by comparing computed energies and transition rates with those obtained through other, more conventional, optimization schemes, as well as with different codes. We compared the results with calculations employing the same CSF basis definition but without targeted optimization of the orbitals, thus excluding the influence of added electron correlation. We found that, in comparison to the other optimization schemes we tried, the energies and transition rates obtained with TO showed the best agreement with the experimental values. The improvement in energies was particularly pronounced for the $5d^{10}$nl states. Additionally, the TO method effectively rectified the issue of incorrect relative energy positions in the odd states. We also carried out small-scale calculations with the \textsc{Fac} code and they agree well with \textsc{Grasp} when targeted optimization is used, but with TO still yielding a better energy level ordering than \textsc{Fac}.

Certain challenges persist, such as achieving a balance between the optimization of the $5d^{10} nl$ and $5d^{9} nl n'l'$ series simultaneously, especially concerning the even states. We found that while the energy of the $5d^{10} nl$ states improved, the $5d^{9} 6s^{2}$ states were too high relative to the experimental values. Furthermore, for systems such as Au~I that show high mixing between states, in both jj or LSJ representation, the labelling and identification of some levels pose difficulties, thereby complicating comparisons with experimental data as well as other theoretical works.

This work highlights the impact of the choice of orbital optimization scheme and the importance of incorporating non-orthonormal calculations in small-scale calculations that do not rely on a converged energy structure through a systematically expanded CSF basis. While codes such as the (Dirac) B-spline R-matrix, codes by Zatsarinny \citep[(D)BSR;][]{zatsarinny_BSR_2006,zatsarinny_DBSR_2008} allow for fully non-orthonormal structure calculations, they are fundamentally scattering codes and computationally intractable for calculations of complete, spectrum calculations of heavy elements. Implementations of fully non-orthonormal atomic structure methods pose many numerical challenges, making the TO method an effective compromise.

Looking ahead, the TO method is promising for application to more complex systems, such as the lanthanide group of open f-shell elements.
Consider, for instance, Gd I, which shares a similar structure to the copper-group elements that Au is part of, featuring configurations with seven and eight f-electrons and degenerate energy levels from different Rydberg series. We predict significant non-orthonormality effects in for example the 4f and 6s orbitals from the $4f^7$ and $4f^8$ series, which could be addressed through targeted optimization to obtain a more accurate average orbital for these two series. Furthermore, this method can be seen as a correction of small-scale structure models in e.g. multi-element efforts in astrophysical applications, or as a basis for large-scale calculations of single ions targeting high-accuracy atomic data.

\section{Acknowledgements}
We thank Per Jönsson (Malm\"{o} University), Tomas Brage (Lund University) and Alan Hibbert (Queen's University Belfast) for valuable discussions on the atomic structure models and semi-empirical fine-tuning. JG and AMA thank the Swedish Research Council for individual starting grants with contract numbers 2020-05467 and 2020-03940.

\newcommand{\newblock}{}
\bibliography{refs} 

\begin{thebibliography}{49}
\expandafter\ifx\csname natexlab\endcsname\relax\def\natexlab#1{#1}\fi

\bibitem[{Bar-Shalom {et~al.}(2001)Bar-Shalom, Klapisch, \& Oreg}]{hullac_2001}
Bar-Shalom, A., Klapisch, M., \& Oreg, J. 2001,
  \href{http://dx.doi.org/10.1016/S0022-4073(01)00066-8}{\color{blue}Journal of
  Quantitative Spectroscopy and Radiative Transfer}, 71, 169

\bibitem[{Brown \& Ginter(1978)}]{brown_1978}
Brown, C.~M. \& Ginter, M.~L. 1978,
  \href{http://dx.doi.org/10.1364/JOSA.68.000243}{\color{blue}JOSA}, 68, 243

\bibitem[{Böker {et~al.}(2022)Böker, Arribas, Lützgendorf, Alves
  De~Oliveira, Beck, Birkmann, Bunker, Charlot, De~Marchi, Ferruit, Giardino,
  Jakobsen, Kumari, López-Caniego, Maiolino, Manjavacas, Marston, Moseley,
  Muzerolle, Ogle, Pirzkal, Rauscher, Rawle, Rix, Sabbi, Sargent, Sirianni,
  Te~Plate, Valenti, Willott, \& Zeidler}]{boker_2022}
Böker, T., Arribas, S., Lützgendorf, N., {et~al.} 2022,
  \href{http://dx.doi.org/10.1051/0004-6361/202142589}{\color{blue}Astronomy \&
  Astrophysics}, 661, A82

\bibitem[{Carlsson(1988)}]{carlsson_1988}
Carlsson, J. 1988,
  \href{http://dx.doi.org/10.1103/PhysRevA.38.1702}{\color{blue}Physical Review
  A}, 38, 1702

\bibitem[{Civi\u{s} {et~al.}(2011)Civi\u{s}, Matulko\'{a}, Cihelka,
  Kubel\'{i}k, Kawaguchi, \& Chernov}]{civis_2011}
Civi\u{s}, S., Matulko\'{a}, I., Cihelka, J., {et~al.} 2011,
  \href{http://dx.doi.org/10.1088/0953-4075/44/10/105002}{\color{blue}Journal
  of Physics B: Atomic, Molecular and Optical Physics}, 44, 105002

\bibitem[{Cowan {et~al.}(2002)Cowan, Sneden, Burles, Ivans, Beers, Truran,
  Lawler, Primas, Fuller, Pfeiffer, \& Kratz}]{Cowan_2002}
Cowan, J.~J., Sneden, C., Burles, S., {et~al.} 2002,
  \href{http://dx.doi.org/10.1086/340347}{\color{blue}The Astrophysical
  Journal}, 572, 861

\bibitem[{Cowan {et~al.}(2021)Cowan, Sneden, Lawler, Aprahamian, Wiescher,
  Langanke, Martínez-Pinedo, \& Thielemann}]{Cowan_2021}
Cowan, J.~J., Sneden, C., Lawler, J.~E., {et~al.} 2021,
  \href{http://dx.doi.org/10.1103/RevModPhys.93.015002}{\color{blue}Reviews of
  Modern Physics}, 93, 015002

\bibitem[{Ehrhardt \& Davis(1971)}]{ehrhardt_1971}
Ehrhardt, J.~C. \& Davis, S.~P. 1971,
  \href{http://dx.doi.org/10.1364/JOSA.61.001342}{\color{blue}JOSA}, 61, 1342

\bibitem[{Fivet {et~al.}(2006)Fivet, Quinet, Bi\'emont, \& Xu}]{fivet_2006}
Fivet, V., Quinet, P., Bi\'emont, E., \& Xu, H.~L. 2006,
  \href{http://dx.doi.org/10.1088/0953-4075/39/17/015}{\color{blue}Journal of
  Physics B: Atomic, Molecular and Optical Physics}, 39, 3587

\bibitem[{Fontes {et~al.}(2020)Fontes, Fryer, Hungerford, Wollaeger, \&
  Korobkin}]{fontes_2020}
Fontes, C.~J., Fryer, C.~L., Hungerford, A.~L., Wollaeger, R.~T., \& Korobkin,
  O. 2020, \href{http://dx.doi.org/10.1093/mnras/staa485}{\color{blue}Monthly
  Notices of the Royal Astronomical Society}, 493, 4143

\bibitem[{Froese~Fischer(1997)}]{froese_fischer_1997}
Froese~Fischer, C. 1997, Computational {Atomic} {Structure} : {An} {MCHF}
  {Approach} (Routledge)

\bibitem[{Froese~Fischer {et~al.}(2019)Froese~Fischer, Gaigalas, Jönsson, \&
  Biero\'{n}}]{froese_fischer_2019}
Froese~Fischer, C., Gaigalas, G., Jönsson, P., \& Biero\'{n}, J. 2019,
  \href{http://dx.doi.org/10.1016/j.cpc.2018.10.032}{\color{blue}Computer
  Physics Communications}, 237, 184

\bibitem[{Gaarde {et~al.}(1994)Gaarde, Zerne, Caiyan, Zhankui, Larsson, \&
  Svanberg}]{gaarde_1994}
Gaarde, M.~B., Zerne, R., Caiyan, L., {et~al.} 1994,
  \href{http://dx.doi.org/10.1103/physreva.50.209}{\color{blue}Physical Review.
  A, Atomic, Molecular, and Optical Physics}, 50, 209

\bibitem[{George {et~al.}(1988)George, Grays, \& Engleman}]{george_1988}
George, S., Grays, A., \& Engleman, R. 1988,
  \href{http://dx.doi.org/10.1364/JOSAB.5.001500}{\color{blue}JOSA B}, 5, 1500

\bibitem[{Gillanders {et~al.}(2021)Gillanders, McCann, Sim, Smartt, \&
  Ballance}]{gillanders_2021}
Gillanders, J.~H., McCann, M., Sim, S.~A., Smartt, S.~J., \& Ballance, C.~P.
  2021, \href{http://dx.doi.org/10.1093/mnras/stab1861}{\color{blue}Monthly
  Notices of the Royal Astronomical Society}, 506, 3560

\bibitem[{Grevesse {et~al.}(2015)Grevesse, Scott, Asplund, \&
  Sauval}]{Grevesse_2015}
Grevesse, N., Scott, P., Asplund, M., \& Sauval, A.~J. 2015,
  \href{http://dx.doi.org/10.1051/0004-6361/201424111}{\color{blue}Astronomy \&
  Astrophysics}, 573, A27

\bibitem[{Gu(2008)}]{gu_2008}
Gu, M.~F. 2008, \href{http://dx.doi.org/10.1139/P07-197}{\color{blue}Canadian
  Journal of Physics}, 86, 675

\bibitem[{Hannaford {et~al.}(1981)Hannaford, Larkins, \& Lowe}]{hannaford_1981}
Hannaford, P., Larkins, P.~L., \& Lowe, R.~M. 1981,
  \href{http://dx.doi.org/10.1088/0022-3700/14/14/004}{\color{blue}Journal of
  Physics B: Atomic and Molecular Physics}, 14, 2321

\bibitem[{Hibbert(1974)}]{hibbert_1974}
Hibbert, A. 1974,
  \href{http://dx.doi.org/10.1088/0022-3700/7/12/004}{\color{blue}Journal of
  Physics B: Atomic and Molecular Physics}, 7, 1417

\bibitem[{Hibbert(2003)}]{hibbert_2003}
Hibbert, A. 2003,
  \href{http://dx.doi.org/10.1088/0953-4075/36/23/008}{\color{blue}Journal of
  Physics B: Atomic, Molecular and Optical Physics}, 36, 4703

\bibitem[{Jannitti {et~al.}(1979)Jannitti, Cantù, Grisendi, Pettini, \&
  Tozzi}]{jannitti_1979}
Jannitti, E., Cantù, A.~M., Grisendi, T., Pettini, M., \& Tozzi, G.~P. 1979,
  \href{http://dx.doi.org/10.1088/0031-8949/20/2/006}{\color{blue}Physica
  Scripta}, 20, 156

\bibitem[{Jönsson {et~al.}(2023{\natexlab{a}})Jönsson, Gaigalas, Fischer,
  Bieroń, Grant, Brage, Ekman, Godefroid, Grumer, Li, \& Li}]{jonsson_2023}
Jönsson, P., Gaigalas, G., Fischer, C.~F., {et~al.} 2023{\natexlab{a}},
  \href{http://dx.doi.org/10.3390/atoms11040068}{\color{blue}Atoms}, 11

\bibitem[{Jönsson {et~al.}(2023{\natexlab{b}})Jönsson, Godefroid, Gaigalas,
  Ekman, Grumer, Li, Li, Brage, Grant, Bieroń, \& Fischer}]{jonsson_2023b}
Jönsson, P., Godefroid, M., Gaigalas, G., {et~al.} 2023{\natexlab{b}},
  \href{http://dx.doi.org/10.3390/atoms11010007}{\color{blue}Atoms}, 11, 7

\bibitem[{Jönsson {et~al.}(2007)Jönsson, He, Froese~Fischer, \&
  Grant}]{Jonsson_2007}
Jönsson, P., He, X., Froese~Fischer, C., \& Grant, I.~P. 2007,
  \href{http://dx.doi.org/10.1016/j.cpc.2007.06.002}{\color{blue}Computer
  Physics Communications}, 177, 597

\bibitem[{Kasen {et~al.}(2017)Kasen, Metzger, Barnes, Quataert, \&
  Ramirez-Ruiz}]{kasen_2017}
Kasen, D., Metzger, B., Barnes, J., Quataert, E., \& Ramirez-Ruiz, E. 2017,
  \href{http://dx.doi.org/10.1038/nature24453}{\color{blue}Nature}, 551, 80

\bibitem[{Kobayashi {et~al.}(2020)Kobayashi, Karakas, \&
  Lugaro}]{Kobayashi_2020}
Kobayashi, C., Karakas, A.~I., \& Lugaro, M. 2020,
  \href{http://dx.doi.org/10.3847/1538-4357/abae65}{\color{blue}The
  Astrophysical Journal}, 900, 179

\bibitem[{Kramida {et~al.}(2022)Kramida, {Yu.~Ralchenko}, Reader, \& {and NIST
  ASD Team}}]{NIST_ASD}
Kramida, A., {Yu.~Ralchenko}, Reader, J., \& {and NIST ASD Team}. 2022, {NIST
  Atomic Spectra Database (ver. 5.10). [Online]. Available:
  {\href{https://physics.nist.gov/asd}{\tt{https://physics.nist.gov/asd}}}
  [2023, September 11]. National Institute of Standards and Technology,
  Gaithersburg, MD.}

\bibitem[{Li {et~al.}(2023)Li, Gaigalas, Li, Chen, \& Jönsson}]{li_2023}
Li, Y., Gaigalas, G., Li, W., Chen, C., \& Jönsson, P. 2023,
  \href{http://dx.doi.org/10.3390/atoms11040070}{\color{blue}Atoms}, 11, 70

\bibitem[{{LIGO Scientific Collaboration and Virgo Collaboration}
  {et~al.}(2017){LIGO Scientific Collaboration and Virgo Collaboration},
  Abbott, Abbott, \& Abbott}]{Abbott_2017}
{LIGO Scientific Collaboration and Virgo Collaboration}, Abbott, B., Abbott,
  R., \& Abbott, T. 2017,
  \href{http://dx.doi.org/10.1103/PhysRevLett.119.161101}{\color{blue}Physical
  Review Letters}, 119, 161101

\bibitem[{Lundberg {et~al.}(2001)Lundberg, Li, \& Jönsson}]{lundberg_2001}
Lundberg, H., Li, Z.~S., \& Jönsson, P. 2001,
  \href{http://dx.doi.org/10.1103/PhysRevA.63.032505}{\color{blue}Physical
  Review A}, 63, 032505

\bibitem[{Malmqvist(1986)}]{malmqvist_1986}
Malmqvist, P.-{\r{A}}. 1986,
  \href{http://dx.doi.org/10.1002/qua.560300404}{\color{blue}Int. J. Quantum
  Chem.}, 30, 479

\bibitem[{McCann {et~al.}(2022)McCann, Bromley, Loch, \&
  Ballance}]{mccann_2022}
McCann, M., Bromley, S., Loch, S.~D., \& Ballance, C.~P. 2022,
  \href{http://dx.doi.org/10.1093/mnras/stab3285}{\color{blue}Monthly Notices
  of the Royal Astronomical Society}, 509, 4723

\bibitem[{Migdalek(2020)}]{migdalek_2020}
Migdalek, J. 2020,
  \href{http://dx.doi.org/10.1016/j.adt.2020.101324}{\color{blue}Atomic Data
  and Nuclear Data Tables}, 133-134, 101324

\bibitem[{Migdalek \& Garmulewicz(2000)}]{migdalek_2000}
Migdalek, J. \& Garmulewicz, M. 2000,
  \href{http://dx.doi.org/10.1088/0953-4075/33/9/305}{\color{blue}Journal of
  Physics B: Atomic, Molecular and Optical Physics}, 33, 1735

\bibitem[{Papoulia {et~al.}(2019)Papoulia, Ekman, Gaigalas, Godefroid,
  Gustafsson, Hartman, Li, Radžiūtė, Rynkun, Schiffmann, Wang, \&
  Jönsson}]{papoulia_2019}
Papoulia, A., Ekman, J., Gaigalas, G., {et~al.} 2019,
  \href{http://dx.doi.org/10.3390/atoms7040106}{\color{blue}Atoms}, 7, 106

\bibitem[{Parpia {et~al.}(1996)Parpia, Fischer, \& Grant}]{parpia_grasp92_1996}
Parpia, F.~A., Fischer, C.~F., \& Grant, I.~P. 1996,
  \href{http://dx.doi.org/10.1016/0010-4655(95)00136-0}{\color{blue}Computer
  Physics Communications}, 94, 249

\bibitem[{Placco {et~al.}(2015)Placco, Beers, Ivans, Filler, Imig, Roederer,
  Abate, Hansen, Cowan, Frebel, Lawler, Schatz, Sneden, Sobeck, Aoki, Smith, \&
  Bolte}]{Placco_2015}
Placco, V.~M., Beers, T.~C., Ivans, I.~I., {et~al.} 2015,
  \href{http://dx.doi.org/10.1088/0004-637X/812/2/109}{\color{blue}The
  Astrophysical Journal}, 812, 109

\bibitem[{Platt \& Sawyer(1941)}]{platt_1941}
Platt, J.~R. \& Sawyer, R.~A. 1941,
  \href{http://dx.doi.org/10.1103/PhysRev.60.866}{\color{blue}Physical Review},
  60, 866

\bibitem[{Pognan {et~al.}(2023)Pognan, Grumer, Jerkstrand, \&
  Wanajo}]{pognan_2023}
Pognan, Q., Grumer, J., Jerkstrand, A., \& Wanajo, S. 2023,
  \href{http://dx.doi.org/10.1093/mnras/stad3106}{\color{blue}Monthly Notices
  of the Royal Astronomical Society}, 526, 5220

\bibitem[{Roederer {et~al.}(2022)Roederer, Lawler, Hartog, Placco, Surman,
  Beers, Ezzeddine, Frebel, Hansen, Hattori, Holmbeck, \&
  Sakari}]{Roederer_2022}
Roederer, I.~U., Lawler, J.~E., Hartog, E. A.~D., {et~al.} 2022,
  \href{http://dx.doi.org/10.3847/1538-4365/ac5cbc}{\color{blue}The
  Astrophysical Journal Supplement Series}, 260, 27

\bibitem[{Roederer {et~al.}(2012)Roederer, Lawler, Sobeck, Beers, Cowan,
  Frebel, Ivans, Schatz, Sneden, \& Thompson}]{Roederer_2012}
Roederer, I.~U., Lawler, J.~E., Sobeck, J.~S., {et~al.} 2012,
  \href{http://dx.doi.org/10.1088/0067-0049/203/2/27}{\color{blue}The
  Astrophysical Journal Supplement Series}, 203, 27

\bibitem[{Safronova \& Johnson(2004)}]{safronova_2004}
Safronova, U.~I. \& Johnson, W.~R. 2004,
  \href{http://dx.doi.org/10.1103/PhysRevA.69.052511}{\color{blue}Physical
  Review A}, 69, 052511

\bibitem[{Smartt {et~al.}(2017)Smartt, Chen, Jerkstrand, Coughlin, Kankare,
  Sim, Fraser, Inserra, Maguire, Chambers, Huber, Krühler, Leloudas, Magee,
  Shingles, Smith, Young, Tonry, Kotak, Gal-Yam, Lyman, Homan, Agliozzo,
  Anderson, Angus, Ashall, Barbarino, Bauer, Berton, Botticella, Bulla, Bulger,
  Cannizzaro, Cano, Cartier, Cikota, Clark, De~Cia, Della~Valle, Denneau,
  Dennefeld, Dessart, Dimitriadis, Elias-Rosa, Firth, Flewelling, Flörs,
  Franckowiak, Frohmaier, Galbany, González-Gaitán, Greiner, Gromadzki,
  Guelbenzu, Gutiérrez, Hamanowicz, Hanlon, Harmanen, Heintz, Heinze,
  Hernandez, Hodgkin, Hook, Izzo, James, Jonker, Kerzendorf, Klose,
  Kostrzewa-Rutkowska, Kowalski, Kromer, Kuncarayakti, Lawrence, Lowe, Magnier,
  Manulis, Martin-Carrillo, Mattila, McBrien, Müller, Nordin, O’Neill,
  Onori, Palmerio, Pastorello, Patat, Pignata, Podsiadlowski, Pumo, Prentice,
  Rau, Razza, Rest, Reynolds, Roy, Ruiter, Rybicki, Salmon, Schady, Schultz,
  Schweyer, Seitenzahl, Smith, Sollerman, Stalder, Stubbs, Sullivan, Szegedi,
  Taddia, Taubenberger, Terreran, van Soelen, Vos, Wainscoat, Walton, Waters,
  Weiland, Willman, Wiseman, Wright, Wyrzykowski, \& Yaron}]{smartt_2017}
Smartt, S.~J., Chen, T.-W., Jerkstrand, A., {et~al.} 2017,
  \href{http://dx.doi.org/10.1038/nature24303}{\color{blue}Nature}, 551, 75

\bibitem[{Sneden {et~al.}(2003)Sneden, Cowan, Lawler, Ivans, Burles, Beers,
  Primas, Hill, Truran, Fuller, Pfeiffer, \& Kratz}]{Sneden_2003}
Sneden, C., Cowan, J.~J., Lawler, J.~E., {et~al.} 2003,
  \href{http://dx.doi.org/10.1086/375491}{\color{blue}The Astrophysical
  Journal}, 591, 936

\bibitem[{Tanaka {et~al.}(2020)Tanaka, Kato, Gaigalas, \&
  Kawaguchi}]{tanaka_2020}
Tanaka, M., Kato, D., Gaigalas, G., \& Kawaguchi, K. 2020,
  \href{http://dx.doi.org/10.1093/mnras/staa1576}{\color{blue}Monthly Notices
  of the Royal Astronomical Society}, 496, 1369

\bibitem[{Zatsarinny(2006)}]{zatsarinny_BSR_2006}
Zatsarinny, O. 2006,
  \href{http://dx.doi.org/10.1016/j.cpc.2005.10.006}{\color{blue}Computer
  Physics Communications}, 174, 273

\bibitem[{Zatsarinny \& Bartschat(2008)}]{zatsarinny_DBSR_2008}
Zatsarinny, O. \& Bartschat, K. 2008,
  \href{http://dx.doi.org/10.1103/PhysRevA.77.062701}{\color{blue}Physical
  Review A}, 77, 062701

\bibitem[{Zatsarinny \& Froese~Fischer(2016)}]{zatsarinny_2016}
Zatsarinny, O. \& Froese~Fischer, C. 2016,
  \href{http://dx.doi.org/10.1016/j.cpc.2015.12.023}{\color{blue}Computer
  Physics Communications}, 202, 287

\bibitem[{Zhang {et~al.}(2018)Zhang, Zhou, Gao, Yu, Wang, Wang, Gong, \&
  Dai}]{zhang_2018}
Zhang, M., Zhou, L., Gao, Y., {et~al.} 2018,
  \href{http://dx.doi.org/10.1088/1361-6455/aade28}{\color{blue}Journal of
  Physics B: Atomic, Molecular and Optical Physics}, 51, 205001

\end{thebibliography}

\end{document}